\documentclass[%
 aip,
 amsmath,amssymb,
 reprint,%
]{revtex4-1}

\usepackage{graphicx}
\usepackage{dcolumn}
\usepackage{bm}

\usepackage[utf8]{inputenc}
\usepackage[T1]{fontenc}
\usepackage{mathptmx}
\usepackage{etoolbox}

\usepackage{amsmath}
\usepackage{mathtools}
\usepackage{physics}
\usepackage{graphicx}
\usepackage[per-mode=symbol]{siunitx}
\usepackage{ulem}
\DeclareSIUnit\gauss{G}
\DeclareSIUnit\bohr{a_{B}}

\usepackage{hyperref}
\usepackage{bbm}
\usepackage{soul}

\graphicspath{{figures/}}

\usepackage{color}
\definecolor{mygreen}{rgb}{0,0.5,0} 
\definecolor{mygrey}{rgb}{0.5,0.5,0.5} 
\definecolor{myred}{rgb}{0.75,0,0} 
\definecolor{myblue}{rgb}{0,0,0.75} 
\definecolor{mymagenta}{cmyk}{0,1,0,0.12} 
\definecolor{mycyan}{cmyk}{1,0,0,0.12} 
\definecolor{myorange}{rgb}{0.85,0.375,0}  
\definecolor{myviolet}{rgb}{0.5,0.3,1} 
\definecolor{mybrown}{rgb}{0.542969,0.269531, 0.0742188} 

\usepackage{ulem}

\newcommand{\osout}{\bgroup\markoverwith
{\textcolor{myorange}{\rule[0.5ex]{2pt}{0.4pt}}}\ULon}

\makeatletter
\def\@email#1#2{%
 \endgroup
 \patchcmd{\titleblock@produce}
  {\frontmatter@RRAPformat}
  {\frontmatter@RRAPformat{\produce@RRAP{*#1\href{mailto:#2}{#2}}}\frontmatter@RRAPformat}
  {}{}
}%
\makeatother
\begin{document}

\preprint{AIP/123-QED}

\title{Detection of photon-level signals embedded in sunlight with an atomic photodetector}

\newcommand{\ICFO}{ICFO - Institut de Ciencies Fotoniques, The Barcelona Institute of Science and Technology, 08860 Castelldefels, Barcelona, Spain}
\newcommand{\ICREA}{ICREA - Instituci\'{o} Catalana de Recerca i Estudis Avan{\c{c}}ats, 08010 Barcelona, Spain}

\author{Laura Zarraoa}
\email{laura.zarraoa@icfo.eu, morgan.mitchell@icfo.eu}
\affiliation{\ICFO}

\author{Tomas Lamich}%
\affiliation{\ICFO}%

\author{Sondos Elsehimy}
\affiliation{\ICFO}%

\author{Morgan W. Mitchell}
\affiliation{\ICFO}%
\affiliation{\ICREA}%

\author{Romain Veyron}%
\affiliation{\ICFO}%

\date{\today}

\begin{abstract}
The detection of few-photon signals in a broadband background is an extreme challenge for photon counting, requiring filtering that accepts a narrow range of optical frequencies while strongly rejecting all others. Recent work [Zarraoa et. al,  Phys. Rev. Res. 6, 033338 (2024)] demonstrated that trapped single atoms can act as low dark-count narrow-band photodetectors. Here we show that this ``quantum jump photodetector'' (QJPD) approach can also detect photon-level signals embedded in strong sunlight. Using a single rubidium atom as a QJPD, we count arrivals of individual narrow-band laser photons embedded in sunlight powers of order \SI{e10}{photons\per\second}. 
We derive a rate-equation model for the atom's internal-state dynamics in sunlight, and find  quantitative agreement with  experiment.  Using this model, we calculate the channel capacity over a noisy communication channel when sending weak coherent states and detecting them in the presence of sunlight, achieving a representative rate of 0.5 bits per symbol when sending 150 probe photons per \SI{10}{\milli\second}  time-bin, embedded in \SI{1}{\nano\watt} of sunlight  (of order $\SI{e10}{photons\per\second}$ in the visible and near-infrared bands).
The demonstration may benefit background-limited applications such as daytime light detection and ranging (LIDAR), remote magnetometry, and free-space classical and quantum optical communications. 
\end{abstract}

\maketitle

\section{Introduction}
\label{sec:level1}
Detecting weak optical signals embedded in a broadband background such as scattered sunlight is a challenge for many applications. Examples include daylight optical communications \cite{LiaoNatPhot2017, AvesaniNPJqi2021, CaiOptica2024}, daytime photon counting light detection and ranging (LIDAR) \cite{DU2020} and daytime adaptive optics using laser guide stars \cite{HartJATIS2016}. Some strategies to deal with a high  background level include the use of single-mode fibers for spatial filtering \cite{Ntanos2025}, frequency filtering \cite{ZielinskaOL2012, ZielinskaOE2014}, and statistical analysis \cite{Liu2025}. Frequency rejection of a broadband background requires filters with very broad blocking ranges, and with narrow transmission windows around the signal’s frequency. Atoms, due to their strong frequency selectivity, are a natural candidate to build such filters, and have already proven useful as a means of background filtering to increase the signal-to-noise ratios during daytime operation \cite{XiaRS2023, Evans2019}. Despite  advances in background rejection, most space optical communications are still performed during the night to reduce sunlight background. The same happens in other applications such as remote magnetometry using mesospheric sodium \cite{HigbiePNAS2011, KaneJGPSP2018, PedrerosBustosNature2018, Akulshin2025}, currently only performed at night, that would benefit from a detector immune to this background. Daylight operation would provide continuous observation of Earth's magnetic field and shed light on how it interacts with space weather \cite{SpogliSW2023}.

In atmospheric sodium magnetometry, the sunlight background not only affects the detectors but also has an effect on the sodium atoms themselves: it can depolarize the atomic ensemble and lead to decoherence of the spin precession. Besides magnetometry, there is a growing number of applications where understanding the interaction of blackbody radiation (BBR) with matter is becoming important. For instance, room-temperature BBR contributes to frequency shifts in atomic clocks \cite{BeloyPRL2006,HassanPRL2025}, limits the lifetime of Rydberg states\cite{GallagherPRL1979,FarleyPRA1981,BeterovPRA2009} and the states  of molecules\cite{HoekstraPRL2007} due to BBR absorption and stimulated emission, generates attractive forces on atoms \cite{SonnleitnerPRL2013,HaslingerNatPhy2018} and is a source of decoherence of matter waves in heated enclosures \cite{DcampsPRA2024}. 
Fewer works \cite{YounesPRE2024,YounesEntropy2025} have studied the interaction of high-temperature BBR, such as sunlight, with atoms. In this case, the BBR spectrum is distributed around optical frequencies and therefore can directly drive atomic transitions, in contrast to the low BBR temperature which contributes mainly to atomic frequency shifts. 

\begin{figure}
    \centering
    \includegraphics[width=0.48\textwidth]{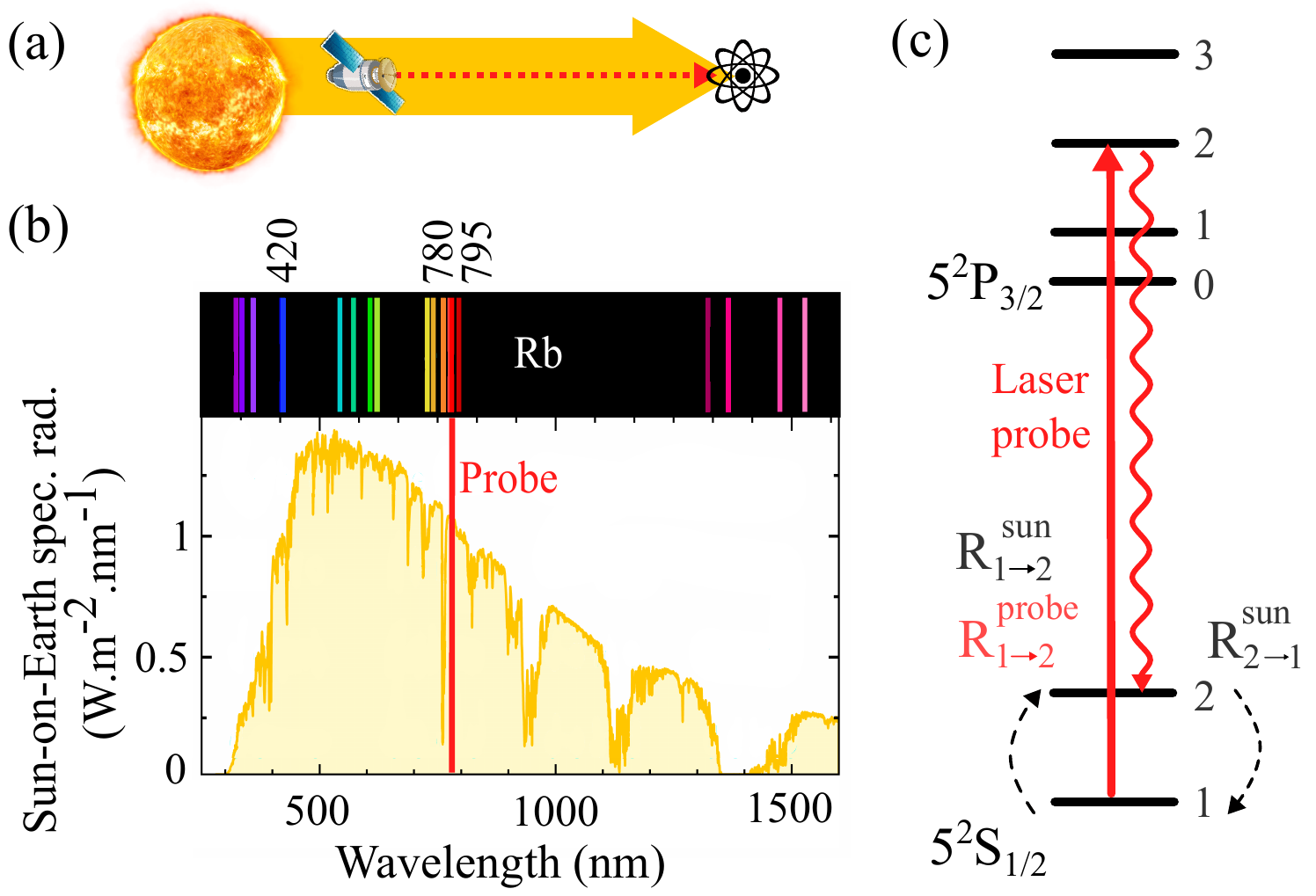}
    \caption{(a) Visual schematic of potential applied scenario: single-photon signals from a satellite are detected using an atomic detector (a single $^{87}$Rb atom) on top of a strong broadband background, e.g., sunlight. (b) Emission lines of $^{87}$Rb with ground-state-connected transitions at 420 nm, 780 nm and 795 nm (top) and sunlight spectrum after passing Earth's atmosphere (bottom). Red solid line marks the probe frequency at \SI{780}{\nano\meter}. (c) Energy levels of $^{87}$Rb with the relevant probe- and sun-driven transition rates.}
    \label{fig:principle}
\end{figure}

In this work, our results are twofold. We provide a quantitative model of sunlight-atom interaction that describes the atom's internal dynamics, and we demonstrate experimentally that a single-atom ``quantum jump photodetector'' (QJPD)\cite{ZarraoaPRR2024}  can detect single-photon signals embedded in sunlight. The paper is organized as follows. Section \ref{sec:setup} presents the single-atom setup used as a QJPD  and describes quantitatively the coupling of sunlight to the atom in the laboratory. Section \ref{sec:theory} provides a model for describing the atom's internal dynamics under both sunlight and weak laser probe illumination. Section \ref{sec:exp} presents experimental results using the QJPD for the detection of signal photons embedded in sunlight, with both initially coupled to a common spatial mode. Its background rejection capabilities are quantified calculating the channel capacity, i.e., the rate at which information can be transmitted over a noisy communication channel.

\section{\label{sec:setup}Experimental setup}
\begin{figure*}
    \centering
\includegraphics[width=\textwidth]{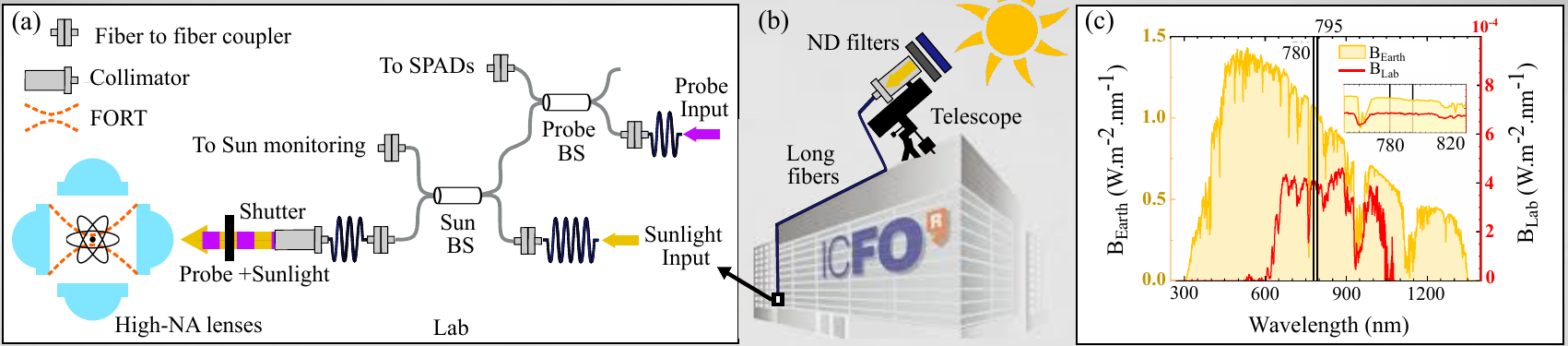}
    \caption{Experimental setup and sunlight spectrum. (a) Sunlight and probe are combined using two fiber beam-splitters (BSs) before being sent to the atom. (b) Sunlight is collected into a single-mode fiber using a collimator mounted on the tracking base of a telescope situated on the roof of the building and sent to the lab via long fibers. (c) Comparison between the spectral radiance of sunlight past the Earth's atmosphere (yellow shaded curve) and the one measured in the lab before the chamber, shown in red. Two black vertical lines at wavelengths \SI{780}{\nano\meter} and \SI{795}{\nano\meter} are included for visual reference. {The inset shows a zoom of the horizontal axis in the region around the wavelengths of interest.}
 }
    \label{fig:Setup}
\end{figure*}

In a previous work \cite{ZarraoaPRR2024}, we demonstrated a single-atom QJPD for the detection of weak coherent light and quantified its characteristics, including quantum efficiency and dark counts.
Our QJPD consists of a single $^{87}$Rb atom trapped by a far-off-resonance trap (FORT) surrounded by four high-NA lenses\cite{BrunoOE2019, BianchetORE2021, ZarraoaPRR2024}. It measures photons by detecting quantum jumps - changes between atomic states with different fluorescence behavior, in our case between the $\mathrm{F}=1$ and $\mathrm{F}=2$ hyperfine levels of the Rb 5\textsuperscript{2}S\textsubscript{1/2} ground state (hereafter $\ket{1}$, and $\ket{2}$, respectively)  - caused by photon absorptions (\autoref{fig:principle}c). 

The experimental system is shown schematically in \autoref{fig:Setup}. The atom is loaded into the FORT, with trap depth  \SI{790}{\micro\kelvin}, cooled by polarization gradient cooling down to \SI{20}{\micro\kelvin} and then optically pumped to 
$\ket{1}$ by a \SI{27}{\milli\second} duration pulse tuned $3\Gamma_0$ to the red of the $F=2 \rightarrow F'=3$ transition, where $\Gamma_0=2\pi \times \SI{6.065}{\mega\hertz}$ is the transition linewidth. It is then exposed to sunlight and probed for a duration of $t_{\text{exp}} = \SI{10}{\milli\second}$. Sunlight can excite the atom via both the $D_1$ and $D_2$ transitions, whereas the probe is tuned to resonance with the $F=1 \rightarrow F'=2$ transition of the D$_2$ line, as seen in \autoref{fig:principle}c. The atomic state is then read out for \SI{1}{\milli\second} by hyperfine-state selective fluorescence detection.
For the experiments presented here, the measured quantum jump efficiency for single photon absorption {on the $F=1 \rightarrow F'=2$ transition} is $\eta_\text{QJ}$ = \SI{8.5+-0.5e-3}{}, a factor of two higher than previous work \cite{ZarraoaPRR2024} 
{due to a smaller FORT beam waist and thus better atom localization and better probe focusing, both of which increase the average light-matter coupling. Both probe and sunlight are linearly polarized along the vertical direction set by gravity, orthogonal to the plane of the four high numerical aperture lenses (NA$=0.5$) and along the FORT polarization and magnetic field offset.} 

Sunlight is coupled into a single-mode fiber (SMF) using a collimator  (Schäfter + Kirchhoff GmbH 60FC-4-A6.2s-02) mounted on a tracking LX200-ACF Meade telescope. The light is brought to the lab via concatenated single-mode fibers (Thorlabs 780HP) of lengths \SI{35}{m} and \SI{25}{m}, with a total transmission at \SI{780}{\nano\meter} of \SI{55}{\percent}. The Sun power coupled into the SMF is monitored with a power meter (Thorlabs S151C) that integrates the power over the entire spectrum with a wavelength-dependent responsivity. {As shown in \autoref{fig:Setup}, the monitored sunlight includes a probe contribution that is negligible, since the sunlight power is on the order of µW while the probe is on the order of fW.} The total irradiance of the Sun, integrated over the whole spectrum, at a distance of one astronomical unit is known as the solar constant \cite{Iqbal1983}, and its average value is \SI{1.361}{\kilo\watt / \meter^2}. Around \SI{20}{\percent} to \SI{30}{\percent} of this is lost in the atmosphere \cite{Wald2018}, leading to the typical value of total solar irradiance at Earth's surface of \SI{1}{\kilo\watt / \meter^2}. From that, there are roughly $ \SI{10}{\milli\watt}$ of Sun power arriving to the collimator, given it has an input area of $\approx \SI{e-5}{\meter^2}$. Of those, on bright days, the maximum total power coupled into the fiber was around  \SI{6}{\micro\watt} at the {collimator} and around \SI{3}{\micro\watt} {in} the lab due to the loss in the long fibers.
Such low collection efficiency is expected due to the multi-spatial-mode character of sunlight, of which only one mode can be carried by the fiber \cite{YounesPRE2024}. We note that even for starlight, which has much greater spatial coherence, the efficiency of coupling into SMF can be low, e.g.,  0.45$\%$ in Sliski et al. \cite{Sliski2023}. 

Sunlight and probe are combined and monitored using two fiber 50:50 beamsplitters (BSs, Thorlabs PN780R5A2). The "probe BS" splits the probe between monitoring in single-photon avalanche detectors (SPADs) and the "Sun BS" where it is combined with sunlight. The "Sun BS" splits the combined light between monitoring and the atom chamber. The power of the probe light was controlled by an acousto-optic modulator before reaching the beamsplitters. Different sunlight powers are obtained by combinations of three absorptive neutral density (ND) filters (two NE10A and one NE05A, Thorlabs) placed before the Sun-coupling collimator. The combined light was time-gated with a free-space shutter and focused on the atom using one of the high-NA lenses.

{We directly measured the spectral radiance in the lab before focusing (right axis on \autoref{fig:Setup}c), from which we will later compute the transition rates at the atom position. 
The integrated maximum Sun power {after the "Sun BS" and the optics}, deconvolved from the detector responsivity, is \SI{1.38}{\micro\watt}, which corresponds to \SI{51.2}{\femto\watt} {within the \SI{6}{\mega\hertz} atom linewidth at \SI{780}{\nano\meter}.}}

\section{\label{sec:theory} Model for an atom interacting with a weak laser probe and sunlight} 

In this section, we use rate equations to describe the atom's internal dynamics driven by resonant photons from the probe and sunlight that we later compare to the experimental results. We also provide a quantitative analysis of the effect of off-resonant photons by calculating AC Stark shifts due to sunlight.

\subsection{\label{subsec:level21}Rate equation with laser probe and sunlight}
Sunlight contributes to driving the atomic transitions over a broad spectrum. This results in net transition rates between $\ket{1}$ and $\ket{2}$ as shown in \autoref{fig:principle}. The population $N_2$ of the $\ket{2}$ state evolves under the influence of incoherent light or a weak coherent laser driving in $\ket{2}$ as described by the rate equation
\begin{equation}
    \begin{aligned}
        \frac{dN_2}{dt} = R_{1 \rightarrow 2} N_1 - R_{2 \rightarrow 1} N_2.
        \label{eq:population_dynamics}
    \end{aligned}
\end{equation}
If the atom starts in the state $\ket{1}$, the time evolution of the population $N_2$ is
\begin{equation}
    \begin{aligned}
        N_2(t) = \frac{R_{1 \rightarrow 2}}{R_{1 \rightarrow 2}+R_{2 \rightarrow 1}} \left( 1-e^{-(R_{1 \rightarrow 2}+R_{2 \rightarrow 1})t} \right).
        \label{eq:population_dynamics_solution}
    \end{aligned}
\end{equation}

The transition rates $R_{i \rightarrow j}$ under incoherent sunlight excitation are calculated using the absorption rates given by the Einstein coefficients \cite{FarleyPRA1981,YounesPRE2024}. 
{We neglect stimulated emission whose magnitude can be upper-bounded by Einstein's arguments: If the atom were exposed to omnidirectional BBR at temperature $T$, the ratio of stimulated emission to spontaneous emission rates would be given by the average photon number per angular frequency mode $\bar{n}(\omega)=1/(e^{\hbar\omega/k_bT}-1)$.  For a mode at \SI{780}{\nano\meter} and $T=\SI{5800}{\kelvin}$, $\bar{n}=0.04\ll1$. In our scenario, where the sunlight is inefficiently collected and the illumination on the atom is over a solid angle $\approx \pi/6$, the rate of stimulated emission will be considerably less.} 
The total rate $R_{F_{i} \rightarrow F_{f}}$ from $F_{i}$ to $F_{f}$ is {then} given by an absorption from the initial state $\ket{J_g,F_{i}}$ towards any excited state $\ket{e}=\ket{J_e,F_e}$ with oscillator strength $f_{F_i,F_e}$, followed by spontaneous emission at a rate $\Gamma_{J_eJ_g}$ towards the final state $\ket{J_g,F_{f}}$ with oscillator strength $f_{F_e,F_f}$. {As the atom is considered unpolarized, the oscillator strengths between two hyperfine states are summed over the Zeeman states. The total rate is therefore obtained by summing over all excited hyperfine states} 
\begin{equation}
    \begin{aligned}
        R_{F_i \rightarrow F_f}^\text{sun} & = \sum_{\{e\}} \frac{\pi^2 c^3}{\hbar\omega_{J_gJ_e}^3}\rho(\omega_{J_gJ_e})\Gamma_{J_eJ_g} f_{F_i,F_e}  f_{F_e,F_f}   ,
        \label{eq:rate_absorption_spontaneous}
    \end{aligned}
\end{equation}
where $c$ is the speed of light in vacuum, $\hbar$ is the reduced Planck constant, $\omega_{J_gJ_e}$ is the transition angular atomic frequency between the states $J_g$ and $J_e$, and $\rho(\omega)$ is the spectral energy density for blackbody radiation  defined by Planck's law (energy per volume per angular frequency)
\begin{equation}
    \begin{aligned}
        \rho(\omega)=\frac{\hbar}{\pi^2c^3}\frac{\omega^3}{e^{\hbar\omega/k_bT}-1} ,
        \label{eq:BBR}
    \end{aligned}
\end{equation}
{with $\omega$ the angular frequency, $k_b$ the Boltzmann constant and $T$ the blackbody temperature.}

The incident spectral energy density on Earth from the Sun is given by Eq. (\ref{eq:BBR}) multiplied by the solid angle subtended by the Sun as seen from Earth, times the frequency-dependent transmission through the atmosphere. The energy density can be more conveniently expressed in terms of the spectral radiance $B(\lambda)=2\pi\nu^2\rho(\omega)$ (power per area per wavelength) {where $\lambda$ is the wavelength and $\nu$ the frequency corresponding to $\omega$. This spectral radiance is} shown for reference in \autoref{fig:principle}c and has value around the Rb D lines of $B_\text{Earth}(\lambda_{780})\approx 
\SI[per-mode = power]{1}{\watt\per\meter\squared\per\nano\meter}$. {By comparison, the spectral radiance at the fiber output for sunlight power of $P_\text{sun}=\SI{1.38}{\micro\watt}$ is $B_\text{lab}(\lambda_{780})=\SI{4e-4}{\watt\meter^{-2}\nano\meter^{-1}}$, corresponding to $B_\text{at}(\lambda_{780})=\SI{1.5}{k\watt\meter^{-2}\nano\meter^{-1}}$ in a waist $w_\text{at}=1.3$ \textmu{m} at the atom position.}

In our setup, the part of the Sun spectrum before focusing ranges from \SI{600}{\nano\meter} to \SI{1100}{\nano\meter} (see \autoref{fig:Setup}c). The transitions $5\text{S}_{1/2} \rightarrow 6\text{P}_{1/2}$ at \SI{421}{\nano\meter}  and $5\text{S}_{1/2} \rightarrow 6\text{P}_{3/2}$ at \SI{420}{\nano\meter}, which are present in the Sun radiation, are not transmitted through the fiber and therefore do not contribute to the dynamics. The main transitions that can resonantly interact with Rb atoms are the D$_1$-line at \SI{795}{\nano\meter}  and the D$_2$-line at \SI{780}{\nano\meter}, which overlap with the sunlight spectrum and for which $\rho(\omega_{D_1})\approx\rho(\omega_{D_2})$. 
In addition, the ratios $\Gamma_{J_eJ_g}/\omega_{J_eJ_g}^3$ in Eq. (\ref{eq:rate_absorption_spontaneous}) are equal for the D$_1$- and D$_2$-lines for Rb, which is therefore factorized as a single factor $\Gamma_{0}/\omega_{0}^3$ simplifying the sunlight rates to 

\begin{equation}
    \begin{aligned}
        R_{F_i \rightarrow F_f}^\text{sun} = R_0^\text{sun} \sum_{\{e\}}  f_{F_i,F_e}  f_{F_e,F_f}.
        \label{eq:rates_simplified}
    \end{aligned}
\end{equation}

The scaling rate in Eq. (\ref{eq:rates_simplified}) is $R_0^\text{sun}= {(\pi^2 c^3}/{\hbar\omega_{0}^3)}\rho_0$  with $\rho_0=\rho_\text{at}(\omega_{0})\Gamma_{0}$ being the energy density integrated over the atomic linewidth at the angular atomic frequency $\omega_0=2\pi c/\lambda_0$. It directly links to the Sun intensity at $\omega_0$ as  $\rho_0={I_\text{at}(\omega_{0})}/{c}$, and thus to the power $P_\text{sun}(\omega_0)$ for a Gaussian beam intensity by  $\rho_0={2P_\text{sun}(\omega_{0})}/{\pi w_\text{at}^2c}$ where $w_\text{at}$ is the beam waist at the atom position. 
Finally, $P(\omega_0)$, the solar power within the line at $\omega_0$, is related to the total power $P_\text{sun}$  as  $P_\text{sun}(\omega_{0})=\kappa P_\text{sun}$, where $\kappa$ is defined in terms of the {measured} solar spectral radiance {in the laboratory $B_\text{lab}$}: $\kappa \equiv B_\text{lab}(\lambda_0) \Delta\lambda/\int_0^{\infty} B_\text{lab}(\lambda)d\lambda$ and $\Delta \lambda$ is the transition linewidth in wavelength. The rate $R_0^\text{sun}$ can therefore be expressed linearly with the total Sun power
\begin{equation}
    \begin{aligned}
        R_0^\text{sun} = \frac{2\pi c^2}{\hbar\omega_{0}^3w_\text{at}^2} \kappa P_\text{sun}.
        \label{eq:R0_linear_with_Sun}
    \end{aligned}
\end{equation}
{Equations (\ref{eq:rate_absorption_spontaneous}), (\ref{eq:rates_simplified}) and (\ref{eq:R0_linear_with_Sun}) are only valid for incoherent driving. The probe, driving coherently the atom, can be included in the model using an additional rate as it only drives weakly the atom.} Sending on average $\tilde{n}_{\text{ph}}$ atom-resonant probe photons during the time $t$ with a quantum jump efficiency $\eta_\text{QJ}$ between the states $\ket{1}$ and $\ket{2}$ defines the absorption rate
\begin{equation}
    \begin{aligned}
        R_{1\rightarrow2}^{\text{probe}}\equiv \frac{\eta_{\text{QJ}}\tilde{n}_{\text{ph}}}{t}.
        \label{eq:R_probe_etaQJ}
    \end{aligned}
\end{equation}

The theoretical efficiency $\eta_{\text{QJ}}$ depends on the coupling strength of the atomic transition and the localization strength given by the numerical aperture used to focalize the photons \cite{TeyNJP2009}.  It also depends on the spatiotemporal overlap of the incoming photon with the atom radiation pattern. This efficiency can be precisely calibrated experimentally, incorporating in a single factor these effects\cite{ZarraoaPRR2024}.
The total transition rates in Eq. (\ref{eq:population_dynamics_solution}) with both Sun and probe photons are then $R_{1 \rightarrow 2}=R_{1 \rightarrow 2}^\text{sun}+R_{1 \rightarrow 2}^\text{probe}$ and $R_{2 \rightarrow 1}=R_{2 \rightarrow 1}^\text{sun}$.



\subsection{\label{subsec:level22}AC Stark shifts induced by sunlight}
So far we only considered the influence of the resonant Sun photons on the internal dynamics. However, as the light is tightly focused, we should evaluate the effect of the off-resonant photons that could lead to light shifts. The situation is different from atomic clocks where most of the room-temperature BBR is off-resonant with the relevant atomic transition, which enables  calculating the light shift with an average static field plus small dynamic corrections \cite{FarleyPRA1981,DegenhardtPRA2005,BeloyPRL2006}. For sunlight, the spectrum overlaps with the Rb lines, leading to light shifts of opposite sign. To evaluate the sunlight-induced light shift, we use the full calculations of the light shift using the dynamical polarizabilities \cite{AroraPRA2007}. We compute the total shift by summing the contributions at any angular frequency $\omega$ for a blackbody distribution at the Sun temperature $T=5800$ K for a given frequency-dependent atomic polarizability $\alpha(F,\omega)$ for an atom in a hyperfine state $F$ \cite{FarleyPRA1981}
\begin{equation}
    \begin{aligned}
        \delta E_F = -\frac{1}{4 \epsilon_0} \int_0^\infty \alpha(F,\omega) \rho(\omega)  d\omega.
        \label{eq:light_shift} 
    \end{aligned}
\end{equation}

The energy density $\rho(\omega)$ is scaled such that the total power sent to the atom is equal to $P_\text{sun}$ in a waist of 1.3 \textmu{m}. This includes all losses, from sunlight collection to transmission losses towards the atom. Considering linearly polarized sunlight to calculate the polarizability, and performing the integration of Eq. (\ref{eq:light_shift}) gives $\delta E_{5S_{1/2}, \ket{1}}/P_\text{sun}\approx- 2\pi \hbar \times 30$ Hz/\textmu{W}. The excited state presents Zeeman-dependent shifts, with a maximum shift $\delta E_{5P_{3/2}, \ket{2}}/P_\text{sun}\approx 2\pi \hbar \times 285$ Hz/\textmu{W}  for $m_F=0$. This leads to a maximum differential light shift on the probe transition of $2\pi \hbar \times 315$ Hz/\textmu{W}.
For the Sun powers used in this work, this shift is small compared to the trap depth and atomic linewidth. This comes from the fact that the Rb lines are near the center of the Sun spectrum, experiencing light shifts that globally cancel. For instance, keeping only the wavelengths larger than \SI{800}{\nano\meter} in the BBR leads to $\delta E_{5S_{1/2}, \ket{1}, \lambda>800 \text{nm}}/P_\text{sun}\approx-2\pi \hbar \times530$ Hz/\textmu{W} which is indeed higher than for the full BBR spectrum. Nevertheless, the previous numbers are an over-estimation of the expected shift. In practice, the off-resonant photons will also experience a chromatic focal shift due to the aspherical lens used to focus the light, leading to a frequency-dependent reduction of the intensity at the atom that is expected to reduce the light shifts. Note that considering an isotropic polarization for sunlight only changes slightly the results. In that case, the polarizability is scalar as the vector and tensor part average to zero. The maximum differential light shift is $2\pi \hbar \times 220$ Hz/\textmu{W}, with small variations with the Zeeman states. As a result, for rubidium atoms, the main contribution of sunlight is therefore driving the atomic transitions with resonant photons.

\section{\label{sec:exp} Experimental results}
In this section, we describe the detection of signal photons embedded in sunlight, compare it to the presented model and consider a realistic scenario for a communication link between a satellite and the detector. Finally, using these results, we calculate the channel capacity of a communication channel using such a photodetector.

\subsection{\label{subsec:level32}Coupling of sunlight to a single atom}

The preparation-exposure-readout sequence described in Section \ref{sec:setup} is repeated 200 times for each combination of sunlight and probe photons to extract the probability and uncertainty of the atom being in $\ket{2}$ after the exposure time.  \autoref{fig:Data} shows experimental data of the runs when only sunlight (black rectangles) is sent to the atom during the exposure time. Experimentally, the Sun power is monitored by measuring the total power sent to the atom $P_\text{sun}$. In this case, the population of $\ket{2}$ after exposure with sunlight for a given time, $N^{\text{background}}_2(t)$, can be fitted with Eq. (\ref{eq:population_dynamics_solution}) with two free parameters
\begin{equation}
    \begin{aligned}
        N^{\text{background}}_2(t) = N_2^\text{sat,exp} \left( 1-e^{-b_\text{exp}P_\text{sun}t} \right).
        \label{eq:fit_sun_only}
    \end{aligned}
\end{equation}

The fitted saturated population of $\ket{2}$ is $N_2^\text{sat,exp}=\SI{0.66+-0.03}{}$. The rate to reach this saturation under Sun illumination is found to be $b_\text{exp}= \SI{9 +- 2}{\second^{-1}\nano\watt^{-1}}$. {For the exposure time of 10 ms, this corresponds to a total number of scattered photons from sunlight of 0.09 ph/nW which does not significantly impact the atom temperature}.

By summing over all excited states in  Eq. (\ref{eq:rates_simplified}), we get the predicted transition rates between the two hyperfine ground states  $R_{1 \rightarrow 2}^\text{sun}=(10/9)R_0^\text{sun}$ and $R_{2 \rightarrow 1}^\text{sun}={(2}/{3)}R_0^\text{sun}$. This leads to a saturated population equal to  $N_2^\text{sat,th}={(5}/{8)}=0.625$, consistent with the experimental result. The value $5/8$ reflects the fact that the Sun is driving multiple transitions between hyperfine states with slightly different total transfer rates between $\ket{1}$ and $\ket{2}$ (10/9 compared to 2/3). Consequently, the saturated population is determined by the branching ratios. 

We can calculate the expected rate  as $b_\text{th}=(R_{1\rightarrow2}^\text{sun}+R_{2\rightarrow1}^\text{sun})/P_\text{sun}$ using Eqs. (\ref{eq:rates_simplified}) and (\ref{eq:R0_linear_with_Sun}). Using the experimental spectrum, we calculate the fraction $\kappa \approx\SI{3.7e-8}{}$ of the Sun power at the atomic transition out of the total Sun power, leading to $R_0^\text{sun}/P_\text{sun}\approx \SI{8.6}{\second^{-1}\nano\watt^{-1}}$ and a theoretical rate towards saturation  $b_\text{th}={(16}/{9)}R_0^\text{sun}/P_\text{sun}\approx \SI{15}{\second^{-1}\nano\watt^{-1}}$, a $3\sigma$ difference relative to the observation. The discrepancy is {mainly} explained by the wavelength-dependence of the high-NA lens' focal length, which makes impossible optimal focusing of all the resonant light \cite{TeyNJP2009, ZarraoaPRR2024}. {We estimate this effect by decomposing the total rate as sum of the rates of the D$_1$ and D$_2$ transitions $R_0^\text{sun} = (R_\text{0,D$_1$}^\text{sun} + R_\text{0,D$_2$}^\text{sun})/2$. When sending coherent light at 795 nm, we previously measured a reduction of the quantum jump efficiency \cite{ZarraoaThesis2025} by a factor of 3. Applying that correction to the 795 nm part of sunlight, we get $b_\text{th,corr}=2b_\text{th}/3\approx \SI{10}{\second^{-1}\nano\watt^{-1}}$ which is within the experimental uncertainty.}

\subsection{\label{subsec:level33}Detection of signal photons embedded in sunlight}

\begin{figure}[t!]
    \centering
    \includegraphics[width=0.50\textwidth]{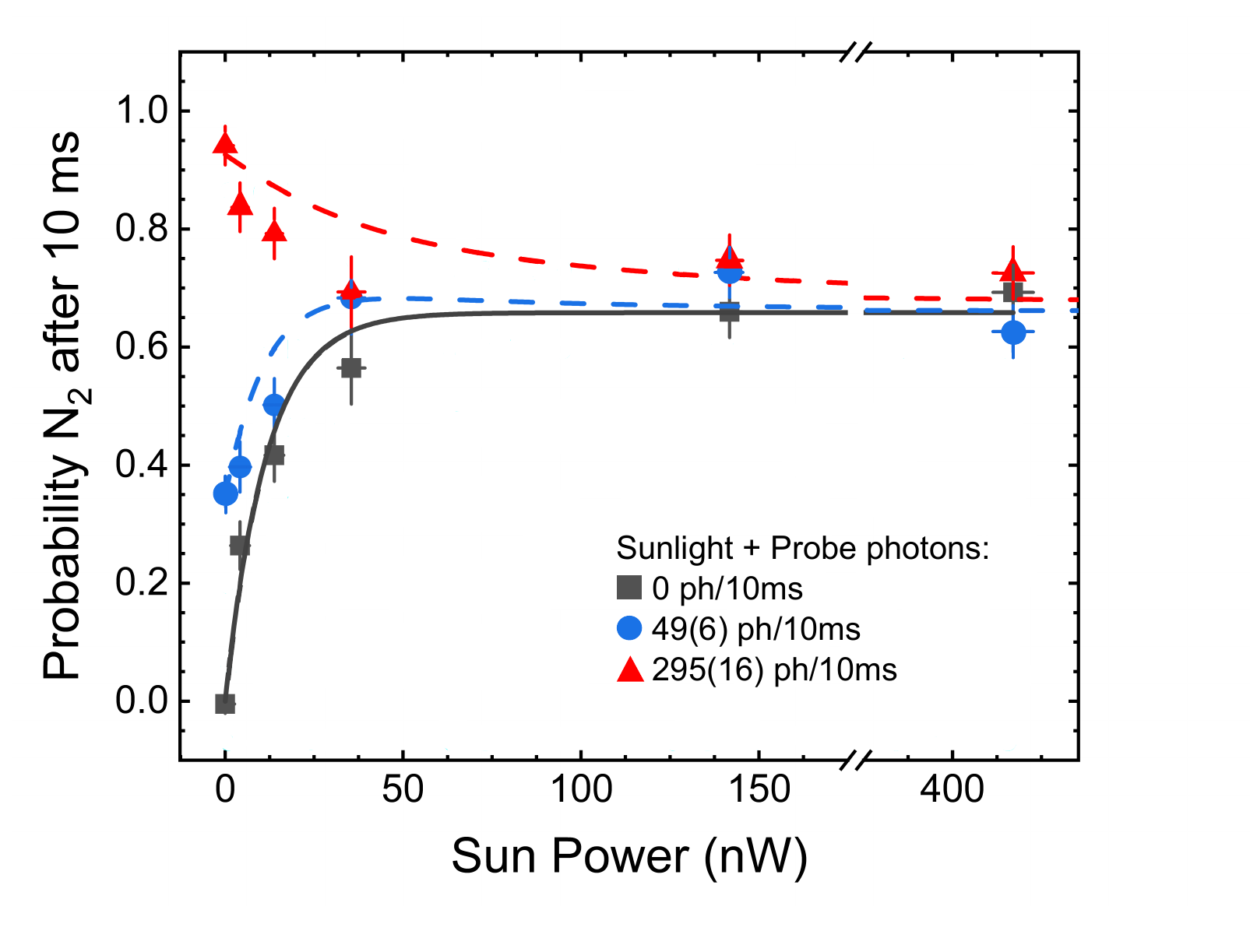}
    \caption{Sunlight combined with probe saturation curves. Data points show the probability of the atom being in $\ket{2}$ after the exposure time of \SI{10}{\milli\second} and its uncertainty, for sunlight combined with probe photons at \SI{0}{ph / 10 \milli \second} (black), \SI{49 +- 6}{ph / 10 \milli \second} (blue) and \SI{295+-16}{ph / 10 \milli \second} (red). {Horizontal error bars show the uncertainty in sun power due to atmospheric disturbance and decoupling during the total measurement time.}
     Solid grey line show data fitted with Eq. (\ref{eq:fit_sun_only}) 
    with parameters $N_2^\text{sat,exp} = \SI{0.66 +- 0.03}{}$ , $b_\text{exp} = \SI{9 +- 2}{\nano \watt^{-1} \second^{-1}}$. Dashed lines plot the theoretical expectation of Eq. (\ref{eq:population_dynamics_solution}) without any free parameters, where the rates were calculated using Eq. (\ref{eq:R_probe_etaQJ}) with $N_2^\text{sat,exp}$ and $b_\text{exp}$ from the fitted Sun data, $\eta_\text{QJ}=\SI{8.5+-0.5e-3}{}$ and corresponding probe photon numbers for low probe (blue) and high probe (red).}
    \label{fig:Data}
\end{figure}

For the rest of the experimental runs, the sunlight was mixed with laser photons tuned to the D\textsubscript{2}, $F = 1 \rightarrow F' = 2$ transition, and the probability to find the atom in $\ket{2}$ under this combined illumination was measured. \autoref{fig:Data} shows experimental points corresponding to sending laser signals of \SI{49 +- 6}{ph / 10~\milli \second} (blue circles) and \SI{295+-16}{ph / 10~\milli \second} (red triangles). {Dashed lines} show the probability of the atom ending in  $\ket{2}$ according to our model given the numbers of probe photons and the fitted Sun parameters $N_2^\text{sat,exp}$ and $b_\text{exp}$, showing a good agreement between experimental data and our model's expectations. 

To give some context for these numbers, we consider an extreme background-rejection problem: a hypothetical space probe flies near to the surface of the Sun, while sending laser signals to an Earth-based telescope in the presence of full solar background, as in \autoref{fig:principle}a. Assuming typical apertures for satellite- and ground-based telescopes \cite{GiggenbachIJSCN2023} of $A_T = \SI{5e-4}{\meter\squared}$ and $A_R = \SI{0.8}{\meter\squared}$ for the transmitter and receiver, respectively, and a link distance of about \SI{1}{AU} (astronomical unit), {the link efficiency is:
\begin{equation}
    \eta_{\text{link}} \approx \frac{4 \pi A_R}{\lambda^2} \left( \frac{\lambda}{4\pi L}\right)^2 \frac{4 \pi A_T}{\lambda^2} = \frac{A_R A_T}{\lambda^2 L^2} \approx  \SI{1e-14}{} .
\end{equation}
}
If the probe emits a \SI{1}{\watt} signal at $\lambda = \SI{780}{\nano\meter}$, the flux of the collected probe photons on Earth would be \SI{400}{photons/10~\milli\second}. This is similar to our $295$ probe photons case for which the signal is well distinguishable from the solar background.

We now calculate the signal-to-noise ratio (SNR) of the QJPD when detecting signal photons mixed with sunlight photons. It is defined using the quantum efficiency and number of sent signal photons, $\eta_{\text{sig}}$ and $N_{\text{sig}}$, and  the quantum efficiency and number of sunlight photons, $\eta_{\text{bg}}$ and $N_{\text{bg}}$
\begin{equation}
    \begin{aligned}
        \text{SNR} = \frac{N_{\text{sig}}\eta_{\text{sig}}}{N_{\text{bg}}\eta_{\text{bg}}}.
        \label{eq:SBRR_SNR}
    \end{aligned}
\end{equation}

The QJPD has detection efficiency on the probe transition $\eta_{\text{sig}}=\eta_{\text{QJ}}$. It responds to sunlight photons within the \SI{6}{\mega\hertz} transition linewidth {of} all transitions causing a jump from $\ket{1}$ to $\ket{2}$. From the total rate given by Eq. (\ref{eq:rates_simplified}), we already calculated $R_{1 \rightarrow 2}^\text{sun}=(10/9)R_0^\text{sun}$, from which the {contribution from the transition between $F=1$ and $F'=2$} is $(5/12)R_0^\text{sun}${, leading to an increase in the quantum jump number by a factor of (10/9)/(5/12)=8/3}. As calculated in the previous section, the fraction of solar photons within the transition linewidth is given by $\kappa \approx\SI{3.7e-8}{}$, leading to a sensitivity to background $\eta_{\text{bg}}=(8/3)\kappa\eta_{\text{QJ}}$. Taking the number of signal photons from the previous example, $N_{\text{sig}}=\SI{400}{ph/10\milli\second}$, and the number of solar photons in \SI{1}{\nano\watt}, $N_{\text{bg}}=\SI{5e7}{ph/10\milli\second}$, the SNR of the QJPD is \SI{80}{}.

For comparison, we consider a filtered detector consisting of a narrowband filter with transmission $T(\lambda)$, maximum transmission $T_\text{max}$, and bandwidth $\Delta \lambda$, used before a broadband detector with quantum efficiency $\eta_{\text{det}}$. In that case, the background sensitivity is $\eta_{\text{bg}}=\eta_{\text{det}}T_{\text{max}} \Delta\lambda  B_\text{lab}(\lambda_0) / \int_0^{\infty} B_\text{lab}(\lambda)d\lambda=\kappa_\text{filter}T_\text{max}\eta_{\text{det}}$ for constant detector efficiency $\eta_{\text{det}}$ across the solar spectrum, while the quantum efficiency for signal is simply $\eta_{\text{sig}}=T_\text{max} \eta_{\text{det}}$.

For example, FADOF filters for Rb have achieved ``ultra-narrow'' bandwidths (equivalent noise bandwidth or ENBW) down to \SI{1}{\giga\hertz} \cite{Zielinska12, Liang2025}, giving $\kappa_\text{filter}=6\times10^{-6}$, with \SI{70}{\percent} maximum transmission. Combining this filter with a detector (e.g., superconducting photon counters) with near-unit detection efficiency would achieve $\eta_{\text{bg}}=4.2\times 10^{-6}$ and $\eta_{\text{sig}}=0.7$. Taking as above $N_{\text{sig}}=\SI{400}{ph/10\milli\second}$ and $N_{\text{bg}}=\SI{5e7}{ph/10\milli\second}$ leads to an SNR equal to \SI{1}{} for this filtered detector, almost two orders of magnitude below the performance of the QJPD detector.  
{Note that the SNR is independent of the detector efficiency and it is mainly determined by the bandwidth of the filtering, which is much narrower for an atom than for a FADOF filter. Higher detector efficiency (or quantum jump efficiency in the case of the QJPD) can however improve other parameters, for example reducing integration times and thus raising transfer rates. } 
\subsection{\label{subsec:level33}Channel capacity}

Let us consider a communication scenario where information is transmitted between a sender (bit $X=0,1$), for example a satellite, and a receiver on Earth (bit $Y=0,1$) in the presence of sunlight background. In information theory, the amount of received information over a classical noisy communication channel can be quantified using Shannon's channel capacity \cite{Shannon1948} that maximizes, over the input probability distribution $q=p(x=1)$, the mutual information $I(X;Y)$ between the sent and received information
\begin{equation}
C  = \text{max}_q~ I(X;Y) ,
        \label{eq:channel_capacity}
\end{equation}
where the mutual information is
\begin{equation}
    \begin{aligned}
        I(X;Y) = \sum_{x,y} p(x) p(y|x) \text{log}_2\frac{p(y|x)}{p(y)}.
        \label{eq:}
    \end{aligned}
\end{equation}

\begin{figure}
    \centering
    \includegraphics[width=0.48\textwidth]{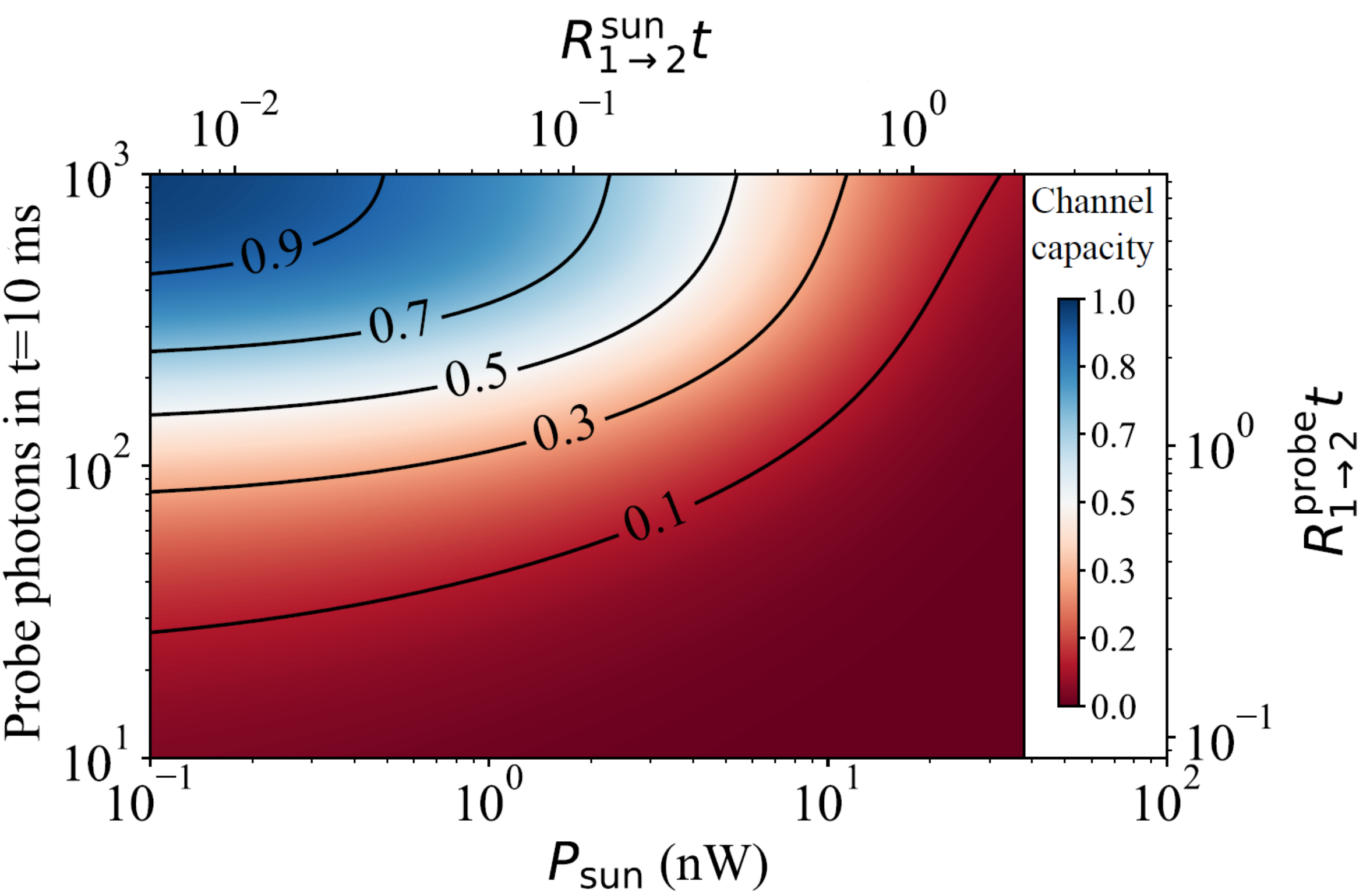}
    \caption{Calculated channel capacity for a binary communication channel in presence of background photons for the atomic detector.}
    \label{fig:channel_capacity}
\end{figure}

We calculate Eq. (\ref{eq:channel_capacity}) using the probability of detecting signal photons $p_s=p(y=1|x=1)=N_2$, and the probability of detecting background photons $p_b=p(y=1|x=0)=N_2^\text{background}$. {In this way, the channel capacity can be calculated as bits per symbol, where the symbol corresponds to the presence (absence) of probe photons for X=1 (X=0)}. \autoref{fig:channel_capacity} shows the correspondent channel capacity. For instance, a channel capacity of 0.5 bits/{symbol} is obtained for 150 probe photons in 10 ms in 1 nW of sunlight. {For those parameters, it represents half of the maximum possible information per channel use,} where error correction codes via redundancy {are typically applied} for transmitting information \cite{Shannon1948}. This motivates the use of an atomic photodetector for low-light detection in broadband background.

\section{\label{sec:level4} Conclusion}
In conclusion, we have demonstrated the background rejection capabilities of using a single atom, in our case $^{87}$Rb, as a photodetector. By measuring quantum jump rates, photon-level signals can be detected with high fidelity even when embedded in up to tens of nW of sunlight power. The atomic dynamics driving the quantum jump rates are modeled using Einstein rate equations to quantify transitions caused by resonant sunlight. The model is found to agree well with experimental data, and is used to further calculate a channel capacity for the detection of optical signals in the presence of sunlight background using a single atom as a photodetector. The AC Stark shift caused by off-resonant background photons is calculated and found to be on the order of a few hundreds of Hz per \textmu{W} of Sun power. This does not significantly contribute to the internal population dynamics for rubidium atoms, but can be {a non negligible effect to account for to adapt this work to narrow-line atoms}.

We studied the extreme case where the background is given by sunlight. However, there are applications dealing with lower background levels, as given by skylight, whose radiance on clear days is {roughly} a factor of 2000 lower in intensity than sunlight \cite{AbbottSmithsonian1932annals}, significantly increasing the channel capacity (\autoref{fig:channel_capacity}).

{We note that through-atmosphere applications, including free-space communications and LIDAR, typically operate with IR or NIR wavelengths, to take advantage of atomospheric transparency. Space communications could in principle benefit from shorter wavelengths to reduce diffraction losses over long links, but in practice mostly work with NIR wavelengths, for which mature laser and detector technologies exist.  Our proposed Rb-based detector at \SI{780}{\nano\meter} thus lies in a preferred wavelength range for these applications. }

{Tuning the source light to match the receiver wavelength is facilitated by the use of an atomic detector -- the same atom could be used by the source as an absolute wavelength reference, with no need for remote calibration. Nonlinear-optical "quantum frequency conversion" can be used to precisely shift the photon frequency \cite{Liao2017,Klop2021, Yang2020, HanADI2023}, either on the source or the detection side, e.g., to compensate for Doppler shifts. AC Stark shifts  \cite{CoopOE2017,Veyron_PRXq_2024} can also be used to shift the receiver's resonance frequency \cite{Ren2022, McSorley2025}. }

{As noted in a previous section, the quantum efficiency does not affect the background rejection capabilities. However, it is directly related to the transfer rate. Ways to improve quantum efficiency and hence achieve higher transfer rates include optically pumping the atom in a specific $m_F$ state to increase the coupling strength \cite{PiroNP2011, Specht2011}, or optical methods including strong beam shaping \cite{SondermannAQT2020}, that increase the overlap of the incoming photon wave-function with the atomic radiation pattern, and cavity enhancement \cite{KuhnPRL2002, NguyenPRA2017} that increases the number of passes of the photon by the atom and thus the chances of interaction. Cavity-QED methods can moreover escape the $\eta_{\textbf{QJ}} \le 1/4$ efficiency limit in driving Raman transitions\cite{Specht2011} such as $F=1\rightarrow F=2$ . } 

{Beyond rate equations, it would be interesting to exploit the coherence of the probe in background rejection since a coherent probe can induce optical atomic coherence whereas an incoherent field cannot.} While we have used the QJPD for detection of classical signals, we note that closely-related techniques have been used for quantum memory and can coherently store quantum information encoded in photon polarization\cite{Specht2011}.

The model for sunlight-atom interaction is general and can be applied to any atom by including the necessary transitions for calculating the rates and the energy shifts. In particular, our results for Rb can be directly applied for studying the internal dynamics of mesospheric sodium \cite{HigbiePNAS2011} which have similar hyperfine structure with D-lines at 589 nm. 

\begin{acknowledgments}
Funded by the European Commission projects Field-SEER (ERC 101097313), OPMMEG (101099379) and QUANTIFY (101135931); Spanish Ministry of Science MCIN project SALVIA (PID2024-158479NB-I00), ``NextGenerationEU/PRTR.'' (Grant FJC2021-047840-I) and ``Severo Ochoa'' Center of Excellence CEX2024-001490-S [MICIU/AEI/10.13039/501100011033];  Generalitat de Catalunya through the CERCA program,  DURSI grant No. 2021 SGR 01453 and QSENSE (GOV/51/2022).  Fundaci\'{o} Privada Cellex; Fundaci\'{o} Mir-Puig. LZ acknowledges the “Presidencia de la Agencia Estatal de Investigación” grant Ref. PRE2020-094392. TL acknowledges Marie Sk\l{}odowska-Curie grant agreement No 847517.  Views and opinions expressed are those of the authors only and do not necessarily reflect those of the European Union or the European Research Council Executive Agency. Neither the European Union nor the granting authority can be held responsible for them.

\end{acknowledgments}

\section*{Data Availability Statement}

The data that support the findings of this study are available from the corresponding author upon reasonable request.

\section*{Author Declarations}

\subsection*{Conflict of interest}
The authors have no conflicts to disclose.

\subsection*{Author Contributions}

\textbf{L. Zarraoa}: Formal Analysis (equal), Investigation (equal), Methodology (equal), Conceptualization (equal),  Writing-Original draft Preparation (equal), Writing-Review and Editing (equal).
\textbf{T. Lamich}: Investigation (equal), Writing-Original draft Preparation (supporting), Writing-Review and Editing (equal).
\textbf{S. Elsehimy}: Investigation (equal), Writing-Original draft Preparation (supporting), Writing-Review and Editing (equal).
\textbf{M.W. Mitchell} : Funding Acquisition (lead), Investigation (equal), Supervision (equal), Conceptualization (equal), Methodology (equal), Writing-Original draft Preparation (equal), Writing-Review and Editing (equal).
\textbf{R. Veyron}: Formal Analysis (equal), Investigation (equal), Methodology (equal), Conceptualization (equal), Supervision (equal), Writing-Original draft Preparation (equal), Writing-Review and Editing (equal).

\section*{References}
\nocite{*}
\bibliography{./biblio/SunAtom}

@article{Holevo_2020,
doi = {10.1070/QEL17285},
url = {https://doi.org/10.1070/QEL17285},
year = {2020},
month = {may},
publisher = {Kvantovaya Elektronika, Turpion Ltd and IOP Publishing},
volume = {50},
number = {5},
pages = {440},
author = {Holevo, A.S.},
title = {Quantum channel capacities},
journal = {Quantum Electronics}
}

@article{Veyron_PRXq_2024,
  title = {In Situ Subwavelength Microscopy of Ultracold Atoms Using Dressed Excited States},
  author = {Veyron, R. and Gerent, J-B. and Baclet, G. and Mancois, V. and Bouyer, P. and Bernon, S.},
  journal = {PRX Quantum},
  volume = {5},
  issue = {3},
  pages = {030349},
  numpages = {16},
  year = {2024},
  month = {Sep},
  publisher = {American Physical Society},
  doi = {10.1103/PRXQuantum.5.030349},
  url = {https://link.aps.org/doi/10.1103/PRXQuantum.5.030349}
}

@article{Zielinska12,
author = {Joanna A. Zieli\'{n}ska and Federica A. Beduini and Nicolas Godbout and Morgan W. Mitchell},
journal = {Opt. Lett.},
number = {4},
pages = {524--526},
publisher = {Optica Publishing Group},
title = {Ultranarrow Faraday rotation filter at the Rb D1 line},
volume = {37},
month = {Feb},
year = {2012},
url = {https://opg.optica.org/ol/abstract.cfm?URI=ol-37-4-524},
doi = {10.1364/OL.37.000524},
}

@article{Shannon1948,
author = {Shannon, C. E.},
title = {A Mathematical Theory of Communication},
journal = {Bell System Technical Journal},
volume = {27},
number = {3},
pages = {379-423},
doi = {https://doi.org/10.1002/j.1538-7305.1948.tb01338.x},
url = {https://onlinelibrary.wiley.com/doi/abs/10.1002/j.1538-7305.1948.tb01338.x},
eprint = {https://onlinelibrary.wiley.com/doi/pdf/10.1002/j.1538-7305.1948.tb01338.x},
year = {1948}
}

@article{HartJATIS2016,
	author = {Michael Hart and Stuart M. Jefferies and Neil Murphy},
	date = {2016-10-26},
	doi = {10.1117/1.jatis.2.4.040501},
	journal = {Journal of Astronomical Telescopes, Instruments, and Systems},
	month = {October},
	number = {4},
	pages = {040501--040501},
	title = {Daylight operation of a sodium laser guide star for adaptive optics wavefront sensing},
	url = {https://lens.org/133-486-071-129-692},
	volume = {2},
	year = {2016}}

@article{Specht2011,
	author = {Specht, Holger P. and Nolleke, Christian and Reiserer, Andreas and Uphoff, Manuel and Figueroa, Eden and Ritter, Stephan and Rempe, Gerhard},
	date = {2011/05/12/print},
	day = {12},
	isbn = {0028-0836},
	journal = {Nature},
	l3 = {http://www.nature.com/nature/journal/v473/n7346/abs/10.1038-nature09997-unlocked.html#supplementary-information},
	m3 = {10.1038/nature09997},
	month = {05},
	number = {7346},
	pages = {190--193},
	publisher = {Nature Publishing Group, a division of Macmillan Publishers Limited. All Rights Reserved.},
	title = {A single-atom quantum memory},
	ty = {JOUR},
	url = {http://dx.doi.org/10.1038/nature09997},
	volume = {473},
	year = {2011}}

@article{CoopOE2017,
	author = {Coop, Simon and Palacios, Silvana and Gomez, Pau and de Escobar, Y. Natali Martinez and Vanderbruggen, Thomas and Mitchell, Morgan W.},
	booktitle = {Optics Express},
	da = {2017/12/25},
	doi = {10.1364/OE.25.032550},
	j2 = {Opt. Express},
	journal = {Optics Express},
	journal1 = {Opt. Express},
	journal2 = {Opt. Express},
	number = {26},
	pages = {32550--32559},
	publisher = {OSA},
	title = {Floquet theory for atomic light-shift engineering with near-resonant polychromatic fields},
	ty = {JOUR},
	url = {http://www.opticsexpress.org/abstract.cfm?URI=oe-25-26-32550},
	volume = {25},
	year = {2017}}

@article{Akulshin2025,
title = {Remote detection optical magnetometry},
journal = {Physics Reports},
volume = {1106},
pages = {1-32},
year = {2025},
note = {Remote Detection Optical Magnetometry},
issn = {0370-1573},
doi = {https://doi.org/10.1016/j.physrep.2024.11.004},
url = {https://www.sciencedirect.com/science/article/pii/S0370157324003855},
author = {Alexander M. Akulshin and Dmitry Budker and Felipe {Pedreros Bustos} and Tong Dang and Emmanuel Klinger and Simon M. Rochester and Arne Wickenbrock and Rui Zhang}
}

@article{ZielinskaOE2014,
	author = {Joanna A. Zieli\'{n}ska and Federica A. Beduini and Vito Giovanni Lucivero and Morgan W. Mitchell},
	doi = {10.1364/OE.22.025307},
	journal = {Opt. Express},
	month = {Oct},
	number = {21},
	pages = {25307--25317},
	publisher = {OSA},
	title = {Atomic filtering for hybrid continuous-variable/discrete-variable quantum optics},
	url = {http://www.opticsexpress.org/abstract.cfm?URI=oe-22-21-25307},
	volume = {22},
	year = {2014}}

@article{ZielinskaOL2012,
	author = {Joanna A. Zieli\'{n}ska and Federica A. Beduini and Nicolas Godbout and Morgan W. Mitchell},
	doi = {10.1364/OL.37.000524},
	journal = {Opt. Lett.},
	month = {Feb},
	number = {4},
	pages = {524--526},
	publisher = {OSA},
	title = {Ultranarrow {F}araday rotation filter at the {R}b {D$_1$} line},
	url = {http://ol.osa.org/abstract.cfm?URI=ol-37-4-524},
	volume = {37},
	year = {2012}}

@article{BrunoOE2019,
	author = {Natalia Bruno and Lorena C. Bianchet and Vindhiya Prakash and Nan Li and Nat\'{a}lia Alves and Morgan W. Mitchell},
	doi = {10.1364/OE.27.031042},
	journal = {Opt. Express},
	month = {Oct},
	number = {21},
	pages = {31042--31052},
	publisher = {OSA},
	title = {Maltese cross coupling to individual cold atoms in free space},
	url = {http://www.opticsexpress.org/abstract.cfm?URI=oe-27-21-31042},
	volume = {27},
	year = {2019}}

@article{BianchetORE2021,
	author = {Bianchet, L C and Alves, N and Zarraoa, L and Bruno, N and Mitchell, M W},
	doi = {10.12688/openreseurope.13972.2},
	journal = {Open Research Europe},
	pages = {102},
	title = {Manipulating and measuring single atoms in the {M}altese cross geometry},
	volume = {1},
	year = {2021}}

@article{SpogliSW2023,
	author = {Spogli, L. and Alfonsi, L. and Cesaroni, C.},
	doi = {https://doi.org/10.1029/2022SW003331},
	eprint = {https://agupubs.onlinelibrary.wiley.com/doi/pdf/10.1029/2022SW003331},
	journal = {Space Weather},
	note = {e2022SW003331 2022SW003331},
	number = {5},
	pages = {e2022SW003331},
	title = {Stepping Into an Equatorial Plasma Bubble With a Swarm Overfly},
	url = {https://agupubs.onlinelibrary.wiley.com/doi/abs/10.1029/2022SW003331},
	volume = {21},
	year = {2023}}

@article{FarleyPRA1981,
  title = {Accurate calculation of dynamic Stark shifts and depopulation rates of Rydberg energy levels induced by blackbody radiation. Hydrogen, helium, and alkali-metal atoms},
  author = {Farley, John W. and Wing, William H.},
  journal = {Phys. Rev. A},
  volume = {23},
  issue = {5},
  pages = {2397--2424},
  numpages = {0},
  year = {1981},
  month = {May},
  publisher = {American Physical Society},
  doi = {10.1103/PhysRevA.23.2397},
  url = {https://link.aps.org/doi/10.1103/PhysRevA.23.2397}
}

@article{BeterovPRA2009,
  title = {Quasiclassical calculations of blackbody-radiation-induced depopulation rates and effective lifetimes of Rydberg $nS$, $nP$, and $nD$ alkali-metal atoms with $n\ensuremath{\le}80$},
  author = {Beterov, I. I. and Ryabtsev, I. I. and Tretyakov, D. B. and Entin, V. M.},
  journal = {Phys. Rev. A},
  volume = {79},
  issue = {5},
  pages = {052504},
  numpages = {11},
  year = {2009},
  month = {May},
  publisher = {American Physical Society},
  doi = {10.1103/PhysRevA.79.052504},
  url = {https://link.aps.org/doi/10.1103/PhysRevA.79.052504}
}

@article{BeloyPRL2006,
  title = {High-Accuracy Calculation of the Blackbody Radiation Shift in the $^{133}\mathrm{Cs}$ Primary Frequency Standard},
  author = {Beloy, K. and Safronova, U. I. and Derevianko, A.},
  journal = {Phys. Rev. Lett.},
  volume = {97},
  issue = {4},
  pages = {040801},
  numpages = {4},
  year = {2006},
  month = {Jul},
  publisher = {American Physical Society},
  doi = {10.1103/PhysRevLett.97.040801},
  url = {https://link.aps.org/doi/10.1103/PhysRevLett.97.040801}
}

@article{HoekstraPRL2007,
  title = {Optical Pumping of Trapped Neutral Molecules by Blackbody Radiation},
  author = {Hoekstra, Steven and Gilijamse, Joop J. and Sartakov, Boris and Vanhaecke, Nicolas and Scharfenberg, Ludwig and van de Meerakker, Sebastiaan Y. T. and Meijer, Gerard},
  journal = {Phys. Rev. Lett.},
  volume = {98},
  issue = {13},
  pages = {133001},
  numpages = {4},
  year = {2007},
  month = {Mar},
  publisher = {American Physical Society},
  doi = {10.1103/PhysRevLett.98.133001},
  url = {https://link.aps.org/doi/10.1103/PhysRevLett.98.133001}
}

@article{YounesPRE2024,
  title = {Laser-type cooling with unfiltered sunlight},
  author = {Younes, Amanda and Campbell, Wesley C.},
  journal = {Phys. Rev. E},
  volume = {109},
  issue = {3},
  pages = {034109},
  numpages = {8},
  year = {2024},
  month = {Mar},
  publisher = {American Physical Society},
  doi = {10.1103/PhysRevE.109.034109},
  url = {https://link.aps.org/doi/10.1103/PhysRevE.109.034109}
}

@Article{YounesEntropy2025,
AUTHOR = {Younes, Amanda and Putnam, Randall and Hamilton, Paul and Campbell, Wesley C.},
TITLE = {Internal State Cooling of an Atom with Thermal Light},
JOURNAL = {Entropy},
VOLUME = {27},
YEAR = {2025},
NUMBER = {3},
ARTICLE-NUMBER = {222},
URL = {https://www.mdpi.com/1099-4300/27/3/222},
PubMedID = {40149146},
ISSN = {1099-4300},
DOI = {10.3390/e27030222}
}

@article{ZarraoaPRR2024,
  title = {Quantum jump photodetector for narrowband photon counting with a single atom},
  author = {Zarraoa, Laura and Veyron, Romain and Lamich, Tomas and Bianchet, Lorena C. and Mitchell, Morgan W.},
  journal = {Phys. Rev. Res.},
  volume = {6},
  issue = {3},
  pages = {033338},
  numpages = {7},
  year = {2024},
  month = {Sep},
  publisher = {American Physical Society},
  doi = {10.1103/PhysRevResearch.6.033338},
  url = {https://link.aps.org/doi/10.1103/PhysRevResearch.6.033338}
}

@article{HaslingerNatPhy2018,
  title = {Attractive force on atoms due to blackbody radiation},
  author = {Haslinger, Philipp and Jaffe, Matt and Xu, Victoria and Schwartz, Osip and Sonnleitner, Matthias and Ritsch-Marte, Monika and Ritsch, Helmut and Müller, Holger},
  journal = {Nature Physics},
  volume = {14},
  issue = {3},
  pages = {257260},
  numpages = {4},
  year = {2018},
  month = {March},
  doi = {10.1038/s41567-017-0004-9},
  url = {https://doi.org/10.1038/s41567-017-0004-9}
}

@article{LiaoNatPhot2017,
  title = {Long-distance free-space quantum key distribution in daylight towards inter-satellite communication},
  volume = {11},
  ISSN = {1749-4893},
  url = {http://dx.doi.org/10.1038/nphoton.2017.116},
  DOI = {10.1038/nphoton.2017.116},
  number = {8},
  journal = {Nature Photonics},
  publisher = {Springer Science and Business Media LLC},
  author = {Liao,  Sheng-Kai and Yong,  Hai-Lin and Liu,  Chang and Shentu,  Guo-Liang and Li,  Dong-Dong and Lin,  Jin and Dai,  Hui and Zhao,  Shuang-Qiang and Li,  Bo and Guan,  Jian-Yu and Chen,  Wei and Gong,  Yun-Hong and Li,  Yang and Lin,  Ze-Hong and Pan,  Ge-Sheng and Pelc,  Jason S. and Fejer,  M. M. and Zhang,  Wen-Zhuo and Liu,  Wei-Yue and Yin,  Juan and Ren,  Ji-Gang and Wang,  Xiang-Bin and Zhang,  Qiang and Peng,  Cheng-Zhi and Pan,  Jian-Wei},
  year = {2017},
  month = jul,
  pages = {509–513}
}

@article{SonnleitnerPRL2013,
  title = {Attractive Optical Forces from Blackbody Radiation},
  author = {Sonnleitner, M. and Ritsch-Marte, M. and Ritsch, H.},
  journal = {Phys. Rev. Lett.},
  volume = {111},
  issue = {2},
  pages = {023601},
  numpages = {5},
  year = {2013},
  month = {Jul},
  publisher = {American Physical Society},
  doi = {10.1103/PhysRevLett.111.023601},
  url = {https://link.aps.org/doi/10.1103/PhysRevLett.111.023601}
}

@article{HigbiePNAS2011,
  title = {Magnetometry with mesospheric sodium},
  volume = {108},
  ISSN = {1091-6490},
  url = {http://dx.doi.org/10.1073/pnas.1013641108},
  DOI = {10.1073/pnas.1013641108},
  number = {9},
  journal = {Proceedings of the National Academy of Sciences},
  publisher = {Proceedings of the National Academy of Sciences},
  author = {Higbie,  James M. and Rochester,  Simon M. and Patton,  Brian and Holzl\"{o}hner,  Ronald and Bonaccini Calia,  Domenico and Budker,  Dmitry},
  year = {2011},
  month = feb,
  pages = {3522–3525}
}

@article{KaneJGPSP2018,
  title = {Laser Remote Magnetometry Using Mesospheric Sodium},
  volume = {123},
  ISSN = {2169-9402},
  url = {http://dx.doi.org/10.1029/2018JA025178},
  DOI = {10.1029/2018ja025178},
  number = {8},
  journal = {Journal of Geophysical Research: Space Physics},
  publisher = {American Geophysical Union (AGU)},
  author = {Kane,  Thomas J. and Hillman,  Paul D. and Denman,  Craig A. and Hart,  Michael and Phillip Scott,  R. and Purucker,  Michael E. and Potashnik,  S. J.},
  year = {2018},
  month = aug,
  pages = {6171–6188}
}

@article{PedrerosBustosNature2018,
  title = {Remote sensing of geomagnetic fields and atomic collisions in the mesosphere},
  volume = {9},
  ISSN = {2041-1723},
  url = {http://dx.doi.org/10.1038/s41467-018-06396-7},
  DOI = {10.1038/s41467-018-06396-7},
  number = {1},
  journal = {Nature Communications},
  publisher = {Springer Science and Business Media LLC},
  author = {Pedreros Bustos,  Felipe and Bonaccini Calia,  Domenico and Budker,  Dmitry and Centrone,  Mauro and Hellemeier,  Joschua and Hickson,  Paul and Holzl\"{o}hner,  Ronald and Rochester,  Simon},
  year = {2018},
  month = sep 
}

@article{Sliski2023,
author = {Sliski, David H. and Blake, Cullen H. and Eastman, Jason D. and Halverson, Samuel},
title = {Seeing the limited coupling of starlight into single-mode fiber with a small telescope},
journal = {Astronomische Nachrichten},
volume = {344},
number = {5},
pages = {e20220080},
doi = {https://doi.org/10.1002/asna.20220080},
url = {https://onlinelibrary.wiley.com/doi/abs/10.1002/asna.20220080},
eprint = {https://onlinelibrary.wiley.com/doi/pdf/10.1002/asna.20220080},
year = {2023}
}

@article{Shaklan1988,
author = {Stuart Shaklan and Francois Roddier},
journal = {Appl. Opt.},
number = {11},
pages = {2334--2338},
publisher = {Optica Publishing Group},
title = {Coupling starlight into single-mode fiber optics},
volume = {27},
month = {Jun},
year = {1988},
url = {https://opg.optica.org/ao/abstract.cfm?URI=ao-27-11-2334},
doi = {10.1364/AO.27.002334},
}

@article{GallagherPRL1979,
  title = {Interactions of Blackbody Radiation with Atoms},
  volume = {42},
  ISSN = {0031-9007},
  url = {http://dx.doi.org/10.1103/PhysRevLett.42.835},
  DOI = {10.1103/physrevlett.42.835},
  number = {13},
  journal = {Physical Review Letters},
  publisher = {American Physical Society (APS)},
  author = {Gallagher,  T. F. and Cooke,  W. E.},
  year = {1979},
  month = mar,
  pages = {835–839}
}

@incollection{Iqbal1983,
title = {Chapter 3 - THE SOLAR CONSTANT AND ITS SPECTRAL DISTRIBUTION},
editor = {Muhammad Iqbal},
booktitle = {An Introduction to Solar Radiation},
publisher = {Academic Press},
pages = {43-58},
year = {1983},
isbn = {978-0-12-373750-2},
doi = {https://doi.org/10.1016/B978-0-12-373750-2.50008-2},
url = {https://www.sciencedirect.com/science/article/pii/B9780123737502500082},
author = {Muhammad Iqbal}
}

@unpublished{Wald2018,
  TITLE = {{BASICS IN SOLAR RADIATION AT EARTH SURFACE}},
  AUTHOR = {Wald, Lucien},
  URL = {https://minesparis-psl.hal.science/hal-01676634},
  NOTE = {working paper or preprint},
  YEAR = {2018},
  MONTH = Jan,
  HAL_ID = {hal-01676634},
  HAL_VERSION = {v1},
}

@article{Tan2014,
doi = {10.1088/2041-8205/789/1/L10},
url = {https://dx.doi.org/10.1088/2041-8205/789/1/L10},
year = {2014},
month = {jun},
publisher = {The American Astronomical Society},
volume = {789},
number = {1},
pages = {L10},
author = {Tan, P. K. and Yeo, G. H. and Poh, H. S. and Chan, A. H. and Kurtsiefer, C.},
title = {MEASURING TEMPORAL PHOTON BUNCHING IN BLACKBODY RADIATION},
journal = {The Astrophysical Journal Letters}
}

@article{DegenhardtPRA2005,
  title = {Calcium optical frequency standard with ultracold atoms: Approaching $10^{-15}$ relative uncertainty},
  volume = {72},
  ISSN = {1094-1622},
  url = {http://dx.doi.org/10.1103/PhysRevA.72.062111},
  DOI = {10.1103/physreva.72.062111},
  number = {6},
  journal = {Physical Review A},
  publisher = {American Physical Society (APS)},
  author = {Degenhardt,  Carsten and Stoehr,  Hardo and Lisdat,  Christian and Wilpers,  Guido and Schnatz,  Harald and Lipphardt,  Burghard and Nazarova,  Tatiana and Pottie,  Paul-Eric and Sterr,  Uwe and Helmcke,  J\"{u}rgen and Riehle,  Fritz},
  year = {2005},
  month = dec 
}

@article{AroraPRA2007,
  title = {Magic wavelengths for the $np-ns$ transitions in alkali-metal atoms},
  volume = {76},
  ISSN = {1094-1622},
  url = {http://dx.doi.org/10.1103/PhysRevA.76.052509},
  DOI = {10.1103/physreva.76.052509},
  number = {5},
  journal = {Physical Review A},
  publisher = {American Physical Society (APS)},
  author = {Arora,  Bindiya and Safronova,  M. S. and Clark,  Charles W.},
  year = {2007},
  month = nov 
}

@article{DcampsPRA2024,
  title = {Decoherence of a matter wave by blackbody radiation},
  volume = {109},
  ISSN = {2469-9934},
  url = {http://dx.doi.org/10.1103/PhysRevA.109.053306},
  DOI = {10.1103/physreva.109.053306},
  number = {5},
  journal = {Physical Review A},
  publisher = {American Physical Society (APS)},
  author = {Décamps,  B. and Gauguet,  A. and Vigué,  J. and B\"{u}chner,  M.},
  year = {2024},
  month = may 
}

@article{HassanPRL2025,
  title = {Cryogenic Optical Lattice Clock with 
$1.7\times10^{-20}$ Blackbody Radiation Stark Uncertainty},
  volume = {135},
  ISSN = {1079-7114},
  url = {http://dx.doi.org/10.1103/4tky-jmsm},
  DOI = {10.1103/4tky-jmsm},
  number = {6},
  journal = {Physical Review Letters},
  publisher = {American Physical Society (APS)},
  author = {Hassan,  Youssef S. and Beloy,  Kyle and Siegel,  Jacob L. and Kobayashi,  Takumi and Swiler,  Eric and Grogan,  Tanner and Brown,  Roger C. and Rojo,  Tristan and Bothwell,  Tobias and Hunt,  Benjamin D. and Halaoui,  Adam and Ludlow,  Andrew D.},
  year = {2025},
  month = aug 
}

@article{AvesaniNPJqi2021,
  title = {Full daylight quantum-key-distribution at 1550 nm enabled by integrated silicon photonics},
  volume = {7},
  ISSN = {2056-6387},
  url = {http://dx.doi.org/10.1038/s41534-021-00421-2},
  DOI = {10.1038/s41534-021-00421-2},
  number = {1},
  journal = {npj Quantum Information},
  publisher = {Springer Science and Business Media LLC},
  author = {Avesani,  M. and Calderaro,  L. and Schiavon,  M. and Stanco,  A. and Agnesi,  C. and Santamato,  A. and Zahidy,  M. and Scriminich,  A. and Foletto,  G. and Contestabile,  G. and Chiesa,  M. and Rotta,  D. and Artiglia,  M. and Montanaro,  A. and Romagnoli,  M. and Sorianello,  V. and Vedovato,  F. and Vallone,  G. and Villoresi,  P.},
  year = {2021},
  month = jun 
}

@article{CaiOptica2024,
  title = {Free-space quantum key distribution during daylight and at night},
  volume = {11},
  ISSN = {2334-2536},
  url = {http://dx.doi.org/10.1364/OPTICA.511000},
  DOI = {10.1364/optica.511000},
  number = {5},
  journal = {Optica},
  publisher = {Optica Publishing Group},
  author = {Cai,  Wen-Qi and Li,  Yang and Li,  Bo and Ren,  Ji-Gang and Liao,  Sheng-Kai and Cao,  Yuan and Zhang,  Liang and Yang,  Meng and Wu,  Jin-Cai and Li,  Yu-Huai and Liu,  Wei-Yue and Yin,  Juan and Wang,  Chao-Ze and Luo,  Wen-Bin and Jin,  Biao and Lv,  Chao-Lin and Li,  Hao and You,  Lixing and Shu,  Rong and Pan,  Ge-Sheng and Zhang,  Qiang and Liu,  Nai-Le and Wang,  Xiang-Bin and Wang,  Jian-Yu and Peng,  Cheng-Zhi and Pan,  Jian-Wei},
  year = {2024},
  month = may,
  pages = {647}
}

@article{TeyNJP2009,
  title = {Interfacing light and single atoms with a lens},
  volume = {11},
  ISSN = {1367-2630},
  url = {http://dx.doi.org/10.1088/1367-2630/11/4/043011},
  DOI = {10.1088/1367-2630/11/4/043011},
  number = {4},
  journal = {New Journal of Physics},
  publisher = {IOP Publishing},
  author = {Tey,  Meng Khoon and Maslennikov,  Gleb and C H Liew,  Timothy and Aljunid,  Syed Abdullah and Huber,  Florian and Chng,  Brenda and Chen,  Zilong and Scarani,  Valerio and Kurtsiefer,  Christian},
  year = {2009},
  month = apr,
  pages = {043011}
}

@Article{Du2020,
AUTHOR = {Du, L. and Wang, J. and Yang, Y. and Xun, Y. and Li, F. and Wu, F. and Gong, S. and Zheng, H. and Cheng, X. and Yang, G. and Lu, Z.},
TITLE = {Continuous Detection of Diurnal Sodium Fluorescent Lidar over Beijing in China},
JOURNAL = {Atmosphere},
VOLUME = {11},
YEAR = {2020},
NUMBER = {1},
ARTICLE-NUMBER = {118},

ISSN = {2073-4433},
DOI = {10.3390/atmos11010118}
}

@article{Jiang2023,
title = {Photon counting lidar working in daylight},
journal = {Optics and Laser Technology},
volume = {163},
pages = {109374},
year = {2023},
issn = {0030-3992},
doi = {10.1016/j.optlastec.2023.109374},
author = {Y. Jiang and B. Liu and R. Wang and Z. Li and Z. Chen and B. Zhao and G. Guo and W. Fan and F. Huang and Y. Yang}
}

@article{XiaRS2023,
	article-number = {5140},
	author = {Xia, Y. and Cheng, X. and Wang, Z. and Liu, L. and Yang, Y. and Du, L. and Jiao, J. and Wang, J. and Zheng, H. and Li, Y. and Li, F. and Yang, G.},
	doi = {10.3390/rs15215140},
	issn = {2072-4292},
	journal = {Remote Sensing},
	number = {21},
	title = {Design of a Data Acquisition, Correction and Retrieval of {Na} {Doppler} Lidar for Diurnal Measurement of Temperature and Wind in the Mesosphere and Lower Thermosphere Region},
	volume = {15},
	year = {2023},
	}

@INPROCEEDINGS{Evans2019,
  author={Evans, Christopher C. and Woolf, David N. and Brown, Justin M. and Hensley, Joel M.},
  booktitle={2019 Conference on Lasers and Electro-Optics (CLEO)}, 
  title={A Daytime Free-Space Quantum-Optical Link using Atomic-Vapor spectral Filters}, 
  year={2019},
  volume={},
  number={},
  pages={1-2},
  doi={10.1364/CLEO_QELS.2019.FM4C.2}}

@article{Gelbwachs1990,
author = {Jerry A. Gelbwachs},
journal = {Opt. Lett.},
number = {4},
pages = {236--238},
publisher = {Optica Publishing Group},
title = {422.7-nm atomic filter with superior solar background rejection},
volume = {15},
month = {Feb},
year = {1990},
url = {https://opg.optica.org/ol/abstract.cfm?URI=ol-15-4-236},
doi = {10.1364/OL.15.000236},
}

@article{Ntanos2025,
author = "Argiris Ntanos and Aristeidis Stathis and Panagiotis Kourelias and Evridiki Kyriazi and Panagiotis Toumasis and Nikolaos K. Lyras and Nikolaos Makris and Sotirios Tsavdaridis and E. M. Xilouris and Athanasios Marousis and Ilias Papastamatiou and Athanasios D. Panagopoulos and Konstantinos Vyrsokinos and Kleomenis Tsiganis and George T. Kanellos and Hercules Avramopoulos and Giannis Giannoulis",
title = "{SMF Coupled Compact Ground Terminal with Advanced Filtering Towards Daylight C Band Satellite QKD}",
year = "2025",
month = "9",
url = "https://preprints.opticaopen.org/articles/preprint/SMF_Coupled_Compact_Ground_Terminal_with_Advanced_Filtering_Towards_Daylight_C_Band_Satellite_QKD/30105127",
doi = "10.48550/arXiv.2509.07667"
}

@article{Liu2025,
author = "Liu, Y. and Dong, X. and Gao, J. and Guan, B. and Zheng, Y. and Liang, Z. and Han, X. and Dong, H.",
title = "{Real-Time Identification Algorithm of Daylight Space Debris Laser Ranging Data Based on Observation Data Distribution Model}",
year = "2025",
journal ="Sensors (Basel, Switzerland)",
volume = "25, 7" ,
month = "4",
url = "https://pmc.ncbi.nlm.nih.gov/articles/PMC11991469/",
doi = "10.3390/s25072281"
}

@article{Xing2025,
    author = "Xing, Ye and Xu, Deifei and Li, Yuan and Zhang, Wuhong and Chen, Lixiang",
    title = "{Sunlight-Excited Spontaneous Parametric Down-Conversion for Quantum Imaging}",
    eprint = "2508.11207",
    archivePrefix = "arXiv",
    primaryClass = "quant-ph",
    month = "8",
    year = "2025"
}

@article{GiggenbachIJSCN2023,
	author = {Giggenbach, D. and Knopp, M. T. and Fuchs, C.},
	doi = {10.1002/sat.1478},
	eprint = {https://onlinelibrary.wiley.com/doi/pdf/10.1002/sat.1478},
	journal = {International Journal of Satellite Communications and Networking},
	number = {5},
	pages = {460-476},
	title = {Link budget calculation in optical LEO satellite downlinks with on/off-keying and large signal divergence: A simplified methodology},
	volume = {41},
	year = {2023},
	}

@article{Liang2025,
author = {Xiaoqian Liang and Yanfeng Bai and Lei Chen and Xiaohui Zhu and Weijun Zhou and Liyu Zhou and Qin Fu and Qi Zhou and Xuanpengfan Zou and Wei Tan and Xianwei Huang and Longfei Yin and Xiquan Fu},
journal = {Opt. Lett.},
number = {13},
pages = {4162--4165},
publisher = {Optica Publishing Group},
title = {Accurate measurement for the orbital angular momentum spectrum in scattering media based on a Faraday atomic filter},
volume = {50},
month = {Jul},
year = {2025},
url = {https://opg.optica.org/ol/abstract.cfm?URI=ol-50-13-4162},
doi = {10.1364/OL.560664},
}

@article{NguyenPRA2017,
  title = {Single atoms coupled to a near-concentric cavity},
  author = {Nguyen, C. H. and Utama, A. N. and Lewty, N. and Durak, K. and Maslennikov, G. and Straupe, S. and Steiner, M. and Kurtsiefer, C.},
  journal = {Phys. Rev. A},
  volume = {96},
  issue = {3},
  pages = {031802},
  numpages = {4},
  year = {2017},
  month = {Sep},
  publisher = {American Physical Society},
  doi = {10.1103/PhysRevA.96.031802}}

@article{KuhnPRL2002,
	author = {Kuhn, A. and Hennrich, M. and Rempe, G.},
	doi = {10.1103/PhysRevLett.89.067901},
	issue = {6},
	journal = {Physical Review Letters},
	journal-iso = {Phys. Rev. Lett.},
	month = {Jul},
	numpages = {4},
	pages = {067901},
	publisher = {American Physical Society},
	title = {Deterministic Single-Photon Source for Distributed Quantum Networking},
	volume = {89},
	year = {2002},
	}

@article{SondermannAQT2020,
author = {Sondermann, M. and Fischer, M. and Leuchs, G.},
title = {Prospects of Trapping Atoms with an Optical Dipole Trap in a Deep Parabolic Mirror for Light–Matter-Interaction Experiments},
journal = {Advanced Quantum Technologies},
volume = {3},
number = {11},
pages = {2000022},
doi = {https://doi.org/10.1002/qute.202000022},
year = {2020}
}

@article{PiroNP2011,
	Author = {Piro, N. and Rohde, F. and Schuck, C. and Almendros, M. and Huwer, J. and Ghosh, J. and Haase, A. and Hennrich, M. and Dubin, F. and Eschner, J.},
	Date = {2011/01//print},
	Isbn = {1745-2473},
	Journal = {Nat Phys},
	M3 = {10.1038/nphys1805},
	Month = {01},
	Number = {1},
	Pages = {17--20},
	Publisher = {Nature Publishing Group},
	Title = {Heralded single-photon absorption by a single atom},
	Ty = {JOUR},
	doi = {10.1038/nphys1805},
	Volume = {7},
	Year = {2011},
	}

@article{Luo2025,
author = {Fubin Luo and Zining Yang and Huizi Zhao and Longfei Jiang and Janyong Sun and Qinshan Liu and Rui Wang and Weiqiang Yang and Hongyan Wang and Xiaojun Xu},
journal = {Opt. Express},
number = {13},
pages = {28832--28840},
publisher = {Optica Publishing Group},
title = {Demonstration of a low-pressure Rb amplifier with an FADOF-coupled ultra-narrowed diode pumping source},
volume = {33},
month = {Jun},
year = {2025},
url = {https://opg.optica.org/oe/abstract.cfm?URI=oe-33-13-28832},
doi = {10.1364/OE.566505},
}

@article{AbbottSmithsonian1932annals,
	author = {Abbot, C. and  Fowle, F. and Aldrich, L.},
	journal = {Annals of the Astrophysical Observatory of the Smithsonian Institution},
	pages = {141},
	publisher = {Smithsonian Astrophysical Observatory, US Government Printing Office},
	title = {The brightness of the sky.},
	volume = {3},
	year = {1932}}

@article{Ren2022,
title = {Adaptive Doppler compensation method for coherent LIDAR based on optical phase-locked loop},
journal = {Measurement},
volume = {187},
pages = {110313},
year = {2022},
issn = {0263-2241},
doi = {https://doi.org/10.1016/j.measurement.2021.110313},
url = {https://www.sciencedirect.com/science/article/pii/S0263224121012112},
author = {Ning Ren and Bin Zhao and Bo Liu and Kangjian Hua}
}

@article{McSorley2025,
  title = {Free-space optical-frequency comparison over rapidly moving links},
  author = {McSorley, Shawn M.P. and Dix-Matthews, Benjamin P. and Frost, Alex M. and McCann, Ayden S. and Karpathakis, Skevos F.E. and Gozzard, David R. and Walsh, Shane M. and Schediwy, Sascha W.},
  journal = {Phys. Rev. Appl.},
  volume = {23},
  issue = {2},
  pages = {L021003},
  numpages = {6},
  year = {2025},
  month = {Feb},
  publisher = {American Physical Society},
  doi = {10.1103/PhysRevApplied.23.L021003},
  url = {https://link.aps.org/doi/10.1103/PhysRevApplied.23.L021003}
}

@Article{Hearne2025,
AUTHOR = {Hearne, Shane and Horgan, Jerry and Boujnah, Noureddine and Kilbane, Deirdre},
TITLE = {Wavelength Selection for Satellite Quantum Key Distribution},
JOURNAL = {Applied Sciences},
VOLUME = {15},
YEAR = {2025},
NUMBER = {3},
ARTICLE-NUMBER = {1308},
URL = {https://www.mdpi.com/2076-3417/15/3/1308},
ISSN = {2076-3417},
DOI = {10.3390/app15031308}
}

@article{HanADI2023,
author = {ZQZ. Han  and ZY. Zhou  and BS. Shi },
title = {Quantum Frequency Transducer and Its Applications},
journal = {Advanced Devices \&amp; Instrumentation},
volume = {4},
number = {},
pages = {0030},
year = {2023},
doi = {10.34133/adi.0030}}

@article{Liao2017,
    author = {Liao, SK. and Yong, HL. and Liu, C. and Shentu, GL. and Li, DD. and Lin, J. and Dai, H. and Zhao, SQ. and Li, B. and Guan, JY. and Chen, W. and Gong, YH. and Li, Y. and Lin, ZH. and Pan, GS. and Pelc, J.S. and Fejer, M.M. and Zhang, WZ. and Liu, WY. and Yin, J. and Ren, JG. and Wang, XB. and Zhang, Q. and Peng, CZ. and Pan, JW.
    } ,
    doi = {10.1038/nphoton.2017.116},
    title = {Long-distance free-space quantum key distribution in daylight towards inter-satellite communication},
    journal = {Nature Photonics} ,
    year = {2017},
    month = {jul}
}

@article{Klop2021,
    author = {Klop, W. and Saathof, R. and Doelman, N. and Gruber, M. and Moens, T. and Osorio Tamayo, C.I. and Duque, C.
    } ,
    doi = {10.1117/12.2599217},
    title = {QKD optical ground terminal developments},
    journal = {International Conference on Space Optics Proceedings} ,
    year = {2021},
    month = {jun}
}

@article{Yang2020,
    author = {Yang, KX. and Abulizi, M. and Li, YH. and Zhang, BY. and Li, SL. and Liu, WY. and Yin, J. and Cao, Y. and Ren, JG. and Peng, CZ.
    } ,
    doi = {10.1364/OE.411939},
    title = {Single-mode fiber coupling with a M-SPGD algorithm for long-range quantum communications},
    journal = {Optics Express} ,
    year = {2020},
    month = {nov}
}

@article{ZarraoaThesis2025,
    author = {Zarraoa, Laura},
    title = {Photon counting with a single neutral atom: quantum efficiency, dark counts, and background rejection},
    journal ={Universitat Politecnica de Catalunya} ,
    volume = {Ph.D. thesis} ,
    year = {2025},
    DOI = {10.5821/dissertation-2117-432886}
}

\end{document}